\documentclass[conference,a4paper]{IEEEtran}
\IEEEoverridecommandlockouts
\usepackage{cite}
\usepackage{amsmath,amssymb,amsfonts}
\usepackage{algorithmic}
\usepackage{graphicx}
\usepackage{textcomp}
\usepackage{xcolor}
\usepackage{caption}
\usepackage{subcaption}
\usepackage{booktabs}
\usepackage{tikz}
\usetikzlibrary{positioning}
\usepackage[pscoord]{eso-pic}
\usepackage{framed}

\usepackage{etoolbox}

\usepackage{url} 

\newcommand{\accessedDate}{April ~2020}
\newcommand{\urlfootnote}[1]{\footnote{\url{#1} Accessed \accessedDate}}

\newcommand{\placetextbox}[3]{
  \AddToShipoutPictureFG*{
    \put(\LenToUnit{#1\paperwidth},\LenToUnit{#2\paperheight}){\vtop{{\null}
        \framebox[\textwidth]{\parbox{\dimexpr\textwidth-2\fboxsep-2\fboxrule}{\footnotesize{#3}}}  
    
    }}%
  }%
}%

\def\BibTeX{{\rm B\kern-.05em{\sc i\kern-.025em b}\kern-.08em
    T\kern-.1667em\lower.7ex\hbox{E}\kern-.125emX}}
\begin{document}

\title{Transformation of Mean Opinion Scores to Avoid Misleading of Ranked based Statistical Techniques}

\author{\IEEEauthorblockN{Babak Naderi}
\IEEEauthorblockA{\textit{Quality and Usability Lab, DFKI} \\
\textit{Technische Universit\"att Berlin}\\
Berlin, Germany \\
babak.naderi@tu-berlin.de}
\and
\IEEEauthorblockN{Sebastian M\"oller}
\IEEEauthorblockA{\textit{Quality and Usability Lab} \\
\textit{Technische Universit\"att Berlin, DFKI}\\
Berlin, Germany \\
sebastian.moeller@tu-berlin.de}
}

\maketitle

\begin{abstract}
The rank correlation coefficients and the ranked-based statistical tests (as a subset of non-parametric techniques) might be misleading when they are applied to subjectively collected opinion scores. 
Those techniques assume that the data is measured at least at an ordinal level and define a sequence of scores to represent a tied rank when they have precisely an equal numeric value.

In this paper, we show that the definition of tied rank, as mentioned above, is not suitable for Mean Opinion Scores (MOS) and might be misleading conclusions of rank-based statistical techniques.
Furthermore, we introduce a method to overcome this issue by transforming the MOS values considering their $95\%$ Confidence Intervals. 
The rank correlation coefficients and ranked-based statistical tests can then be safely applied to the transformed values.
We also provide open-source software packages in different programming languages to utilize the application of our transformation method in the quality of experience domain.

\end{abstract}


\section{Introduction}
\placetextbox{0.07}{0.1}{©2020 IEEE. Personal use of this material is permitted. Permission from IEEE must be obtained for all other uses, in any current or future media, including reprinting/republishing this material for advertising or promotional purposes, creating new collective works, for resale or redistribution to servers or lists, or reuse of any copyrighted component of this work in other works. This paper has been accepted for publication in the 2020 Twelfth International Conference on Quality of Multimedia Experience (QoMEX).}

The ranked based statistical techniques include rank correlation coefficients (i.e. Spearman's and Kendall's rank correlation), and subset of non-parametric groups comparison tests (e.g. Mann-Whitney U Test, Wilcoxon signed-rank test, Kruskal-Wallis Test, Friedman Test).
Those techniques assume that the data is measured at least in an ordinal level.
They define a sequence of scores to represent a \textit{tie rank} when they have exactly an equal numeric value. 
In that case, an equal rank is assigned to all cases in a tie which is the average of the ranks that would have been assigned to each of those cases. 
Sometimes the statistical techniques use a separate formula to calculate their outcome in presence of tied ranks.

The Mean Opinion Score (MOS) represents the average opinion scores. Those scores are the values on a predefined scale that subjects assign to their opinion of the performance of the system \cite{ITU-P10}.
In the listening and/or viewing tests, the most common test procedure is the Absolute Category Rating (ACR) in which the opinion scores are typically collected on a  5-point discrete scale with labels from "Bad" (represented with a numeric value of 1) to "Excellent" (value of 5).
However the resulting MOS value is a continuous number because of the averaging process used to combine opinion scores from different subjects. 
It is recommended to report the subjective MOS with sufficient complementary information about the distribution of ratings i.e. the number of votes and the standard deviation or 95\% Confidence Interval (CI) \cite{ITU-P8002}.
The exact MOS values obtained for a particular condition in a subjective experiment can be influenced by many factors including and not limited to the text instruction, equipment, presentation, preparation of subjects and their profile, and other conditions in the experiment \cite{ITU-P8002}.
Applying the ranked based statistical techniques on the MOS values brings the questions when a set of MOS values should represent a tied rank (or difference between two MOS values should be considered equal in case of repeated measurements). 
In other words, when should two (or more) MOS values be considered to be equal? 

A difference of 0.1 MOS can occur when $90\%$ of subjects gave exactly the same vote to two stimuli, but the scores given by the remaining $10\%$ of subjects deviate 1 point on the ACR scale (e.g. 9 participants vote two stimuli have a "Good" quality and one participant rates first stimulus "Good" and the second on "Excellent"/"Fair"). In the same way, 0.01 MOS difference happens when only $1\%$ of subjects rates with 1 point deviation in ACR scale and all the others give exact same ratings, and for 0.001 MOS deviation, only $0.1\%$ subjects rates with 1 point deviation in ACR scale. 
As a result, the exact same MOS value can rarely be reproduced when repeating the same subjective test, even with the same group of participants.
Meanwhile, software packages use a floating number data type to represent the mean value. Given that, how many fraction digits are making sense for a MOS value?
How tangible a difference between two MOS values is, depends on the number of votes used in the calculation of MOS values.

Given that, it is rare that two stimuli or conditions get an equal MOS value although they represent the same quality i.e. missed tied rank. In this paper, we first show how often such a missed tied rank can happen. Next, we present how missed tied ranks affect the statistical outcome using the Spearman's rank correlation as an example. We also present our transformation method and demonstrate its performance in a simulation study. Finally, we conclude our work and discuss future steps.

\section{Method}

\subsection{Potential tied ranks}

\begin{figure}
    \centering
    \includegraphics[width=\columnwidth]{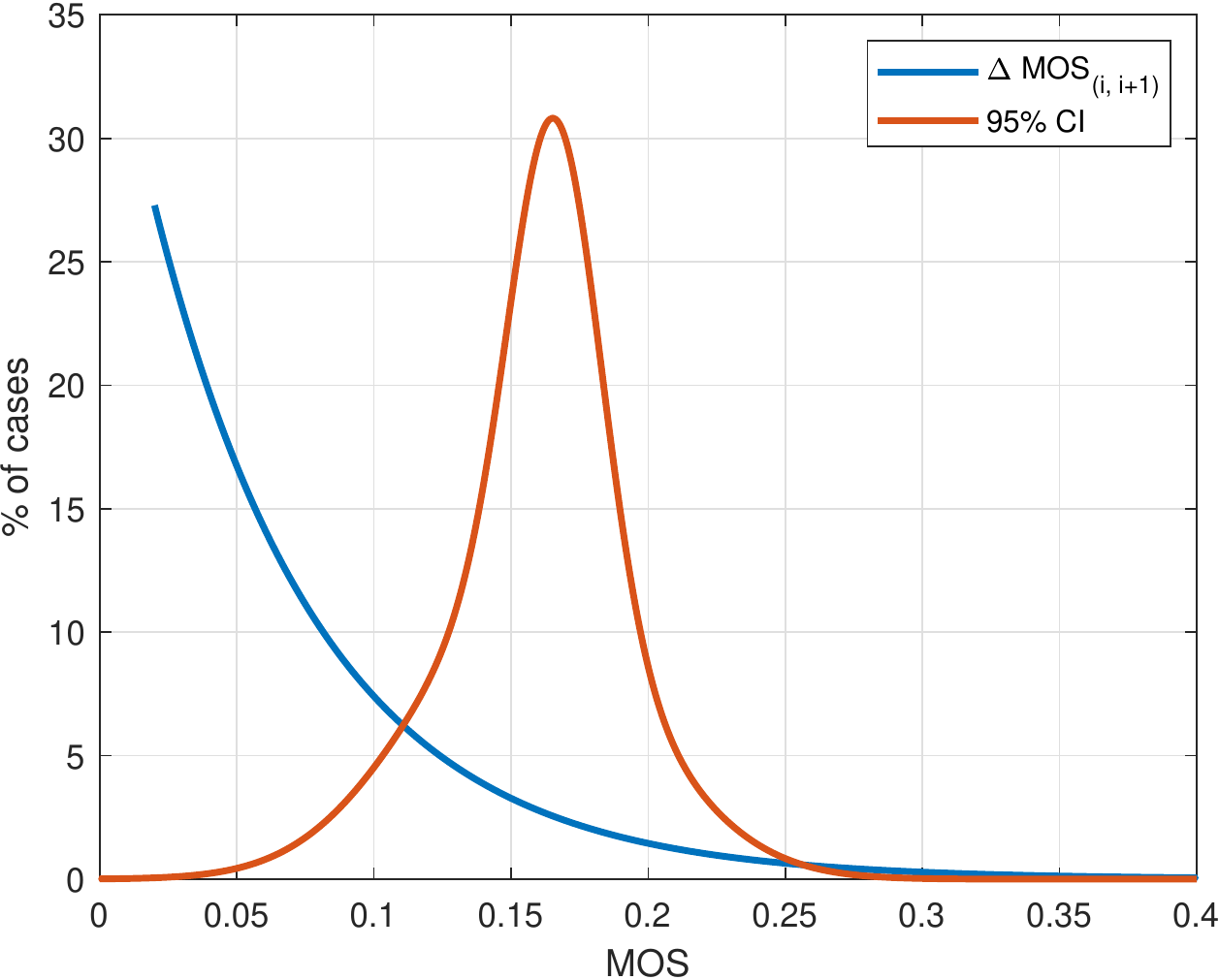}
    \caption{Distribution of 95\% Confidence Intervals and the difference between two MOS values of consecutive conditions in rank order (18 datasets).}
    \label{fig:dist}
\end{figure}

We examined 18 datasets from the domain of speech quality assessment to find out how close are MOS values in a typical dataset and how plausible a tied rank of MOS is.
All datasets were created based on the ITU-T Rec. P.800 \cite{ITU-P800} and included ACR ratings from laboratory-based experiments. 
On average, they contain 50 degradation conditions.
Figure \ref{fig:dist} shows the distribution of 95\%~CIs of the MOS\footnote{The $95\%$~CI is a range of values around the MOS that are believed to contain, with a probability of 95\%, the true MOS value (i.e. the actual population MOS)\cite{field2013discovering}.} and distribution of $\Delta MOS_{i,i+1}$ where $i$ and $i+1$ are two consecutive MOS values in the ranked order from the same dataset. 
In 86\% of cases, the absolute difference between two consecutive MOS values in a ranked order is smaller than the 95\%~CI of one of the conditions. 

\begin{figure}
    \centering
    \includegraphics[width=\columnwidth]{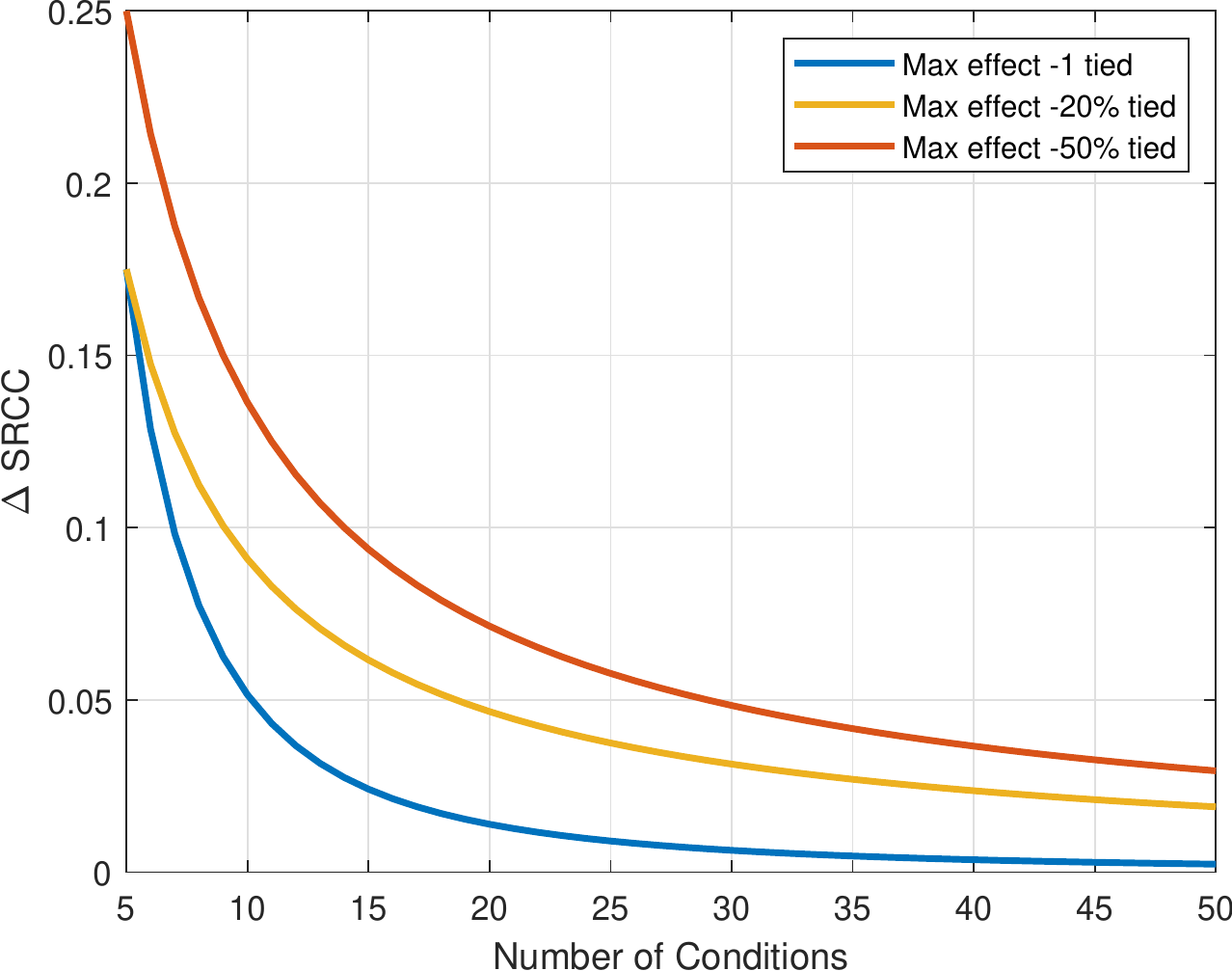}
    \caption{Maximum change in Spearman's ran correlation tied ranks are not recognized.}
    \label{fig:effect}
\end{figure}

For the Spearman's rank correlation, one can calculate the influence of missed tied ranks on the coefficients. $\rho_{A,B}$ reflects the Spearman's rank correlation coefficient between two vectors $A$ and $B$ which can be calculated as follows
\begin{align}
    \rho_{A,B} = 1 - \frac{6 \sum_{i=1}^n d_i^2 }{n(n^2-1)}, && d_{i} = r(A_i)-r(B_i) 
    \label{sc}
\end{align}
where $n$ is the number of conditions, and $r(X_i)$ represent the rank of $i^{th}$ element in vector $X$.
The effect of a missed tie rank on the coefficient can be calculated as following
\begin{align}
    \Delta\rho = |\rho_{A,B} -\rho_{A,B'}|
    \label{d}
\end{align}
where items in $B$ and $B'$ are equal except\footnote{For simplification we only consider two items make a tied rank.} items $i$ and $j$ which get two consecutive rank $k$ and $k+1$ in vector $B$ but are considered to make a tied rank in vector $B'$ and both get rank $k+0.5$.   
Inserting the \eqref{sc} in \eqref{d} leads to
\begin{align}
      \Delta\rho= \bigg|\frac{6}{n(n^2-1)}(d_i^2+d_j^2-d_i'^2-d_j'^2)\bigg| 
      \label{eqn}
\end{align}
\begin{align*}
d_i &= r(A_i)-(k), &   d_j &= r(A_j)-(k+1)\\
d'_i &= r(A_i)-(k+0.5), &   d'_j &= r(A_j)-(k+0.5)
\end{align*}
The maximum difference in coefficient that can result by neglecting $m$ tied ranks can be calculated with the following formula\footnote{Due to the limited space, solving the \eqref{eqn} for maximum effect and generalizing it to $m$ tied ranks, left as an exercise for the reader.}.
\begin{equation}
    \Delta \rho_{max} = \frac{6m(n-m-0.5)  }{n(n^2-1)} 
\end{equation}
where $n$ is the number of conditions.
As a result, when a tied rank is not correctly recognized, it leads to a miss-calculation of the Spearman's rank correlation coefficient (SRCC). 
Figure \ref{fig:effect} illustrates how the effects change based on the number of missed tied ranks and number of conditions. 
The missed tied ranks strongly influence the coefficient when the number of conditions is small ($<$15). 

\subsection{MOS Transformation}
\begin{figure}[ht] 
  \centering
  \begin{subfigure}[b]{0.17\columnwidth}
    \centering
    \includegraphics[width=\textwidth]{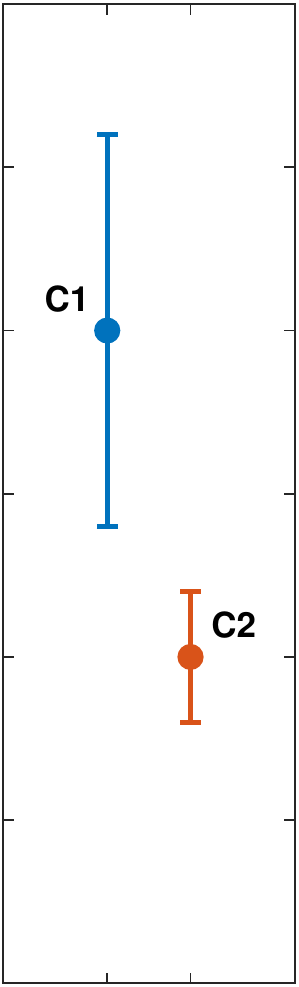} 
    \caption{}
    \label{fig:t_a} 
  \end{subfigure} 
  ~
  \begin{subfigure}[b]{0.17\columnwidth}
    \centering
    \includegraphics[width=\textwidth]{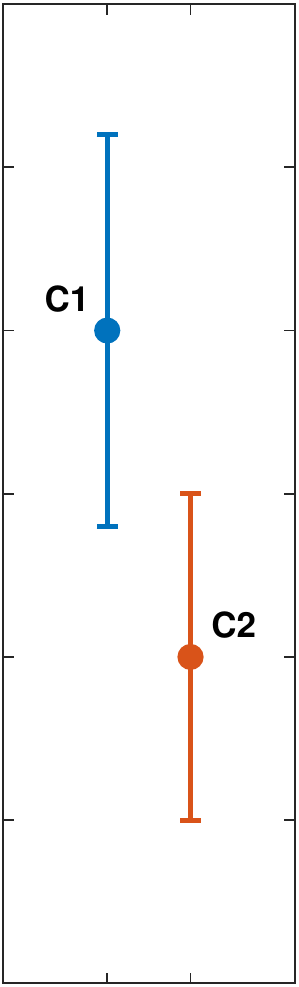}
	\caption{}
	\label{fig:t_b} 
  \end{subfigure} 
  ~
 \begin{subfigure}[b]{0.17\columnwidth}
    \centering
    \includegraphics[width=\textwidth]{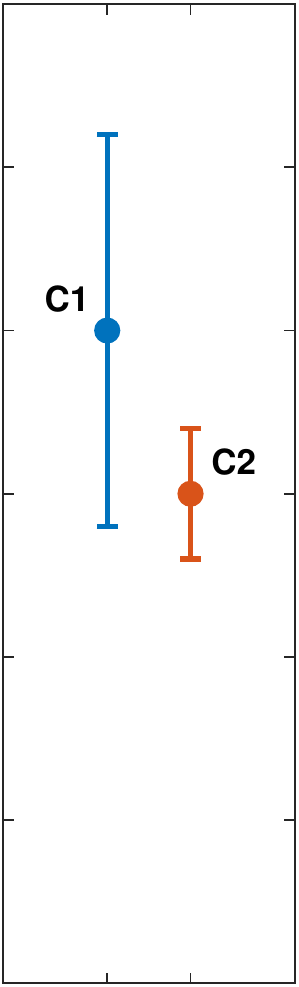}
	\caption{}
	\label{fig:t_c} 
  \end{subfigure} 
  ~
 \begin{subfigure}[b]{0.17\columnwidth}
    \centering
    \includegraphics[width=\textwidth]{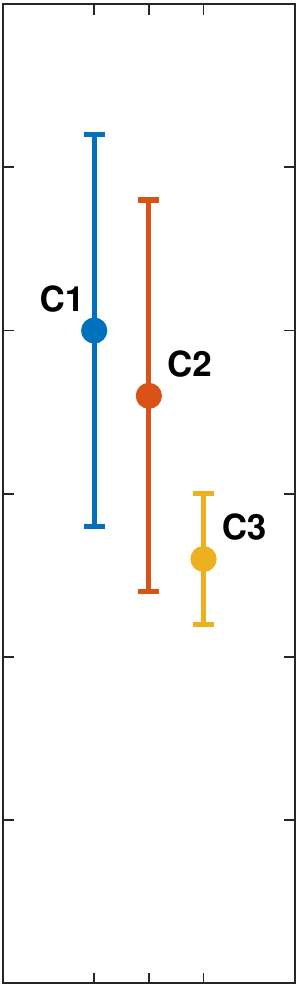}
	\caption{}
	\label{fig:t_d} 
  \end{subfigure} 
  ~
 \begin{subfigure}[b]{0.17\columnwidth}
    \centering
    \includegraphics[width=\textwidth]{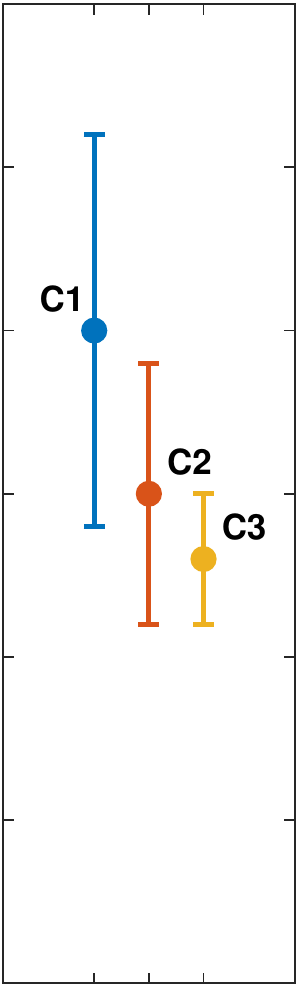}
	\caption{}
	\label{fig:t_e} 
  \end{subfigure} 
  \caption{The transformation method considers C1 and C2 to be a tied rank in (c), and (d). Same as C2 and C3 in (e).}
  \label{fig:t} 
\end{figure}

We define two MOS values represent a tied rank when at least one of them is inside of the 95\%CI of the other one. In that case, two MOS values create a set representing a tied rank. C1 and C2 in Figure \ref{fig:t_c} represents a set of tied rank.
A new MOS value can only be added to an existing set of a tied rank when it makes a tied rank with all members of that set (c.f. Figure \ref{fig:t_d}). Also, when a MOS value (C2) can make a tied rank with two other MOS values (C1, and C3) but not altogether, then it joins to the closest MOS value (i.e. C3 on Figure \ref{fig:t_e}).
In addition, we also round the resulting MOS values to two fraction digits.

We use the above-mentioned rules to transfer the given vectors of MOS values (each accompanied by a vector of 95\%~CIs) to a safe MOS ranked list, which can be directly used in any ranked based statistical technique.
An open-source implementation\urlfootnote{https://github.com/babaknaderi/MOS-transformation} of the transformation method is provided for R, Python, and MATLAB.

\section{Simulation study}
\begin{figure}
    \centering
    \includegraphics[width=\columnwidth]{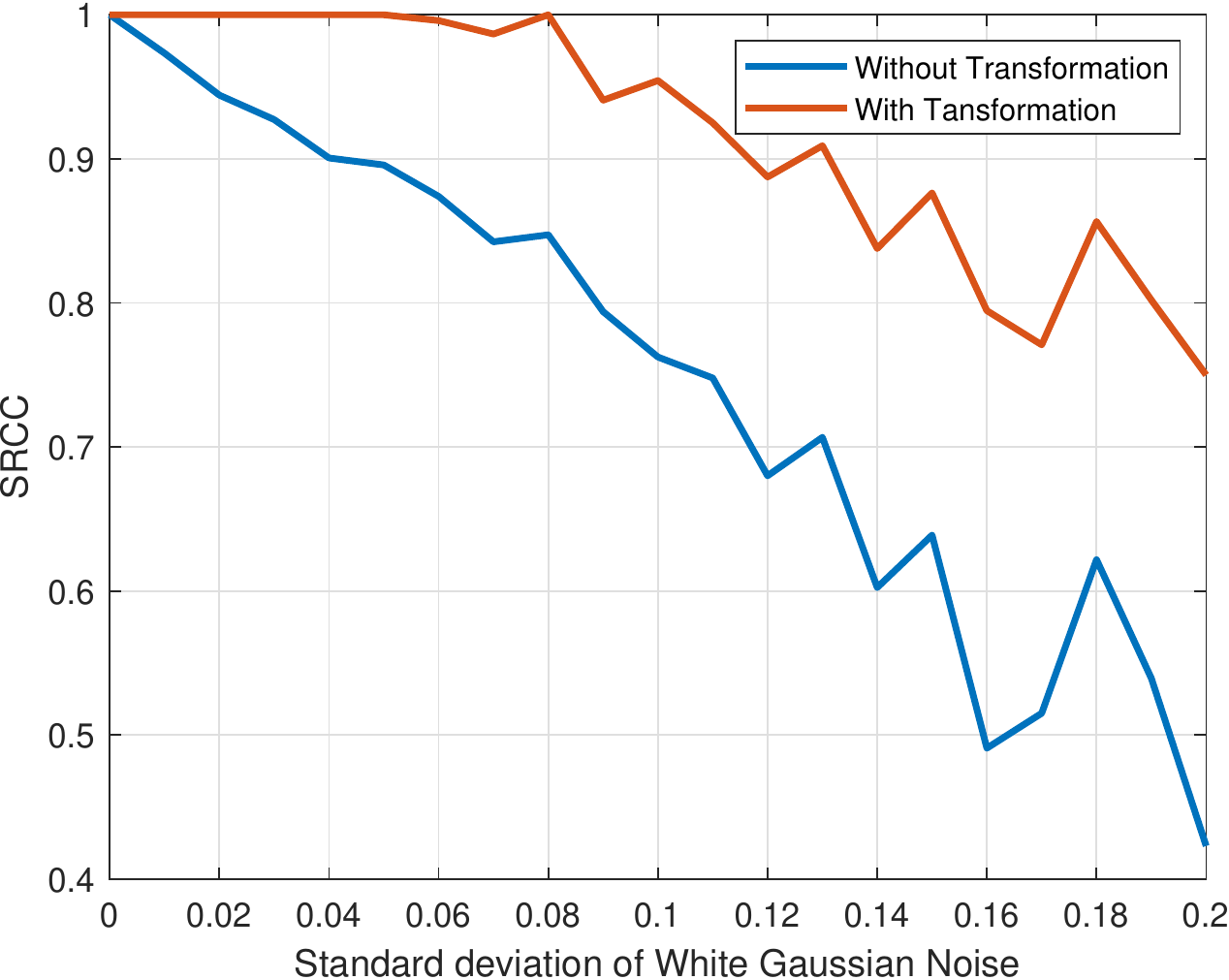}
    \caption{SRCC between true MOS and noisy MOS values with and without implication of the transformation method (largest difference observed in 1000 simulation runs).}
    \label{fig:sim}
\end{figure}

We used a dataset from the speech quality assessment domain with 50 conditions. The subjective ratings were collected in a laboratory test according to the ITU-T Rec. 800 \cite{ITU-P800}. The MOS values, and 95\%~CI were kindly provided to us by the dataset owner.
We took ten conditions with the highest MOS values for the simulation as the effect is more tangible with a small number of conditions.
During the simulation, we took the MOS values (from now on true MOS) and added noise generated by different White Gaussian Noise functions to them (from now on noisy MOS) and calculate the SRCC between the true MOS and noisy MOS with and without our transformation method and their absolute difference ($\Delta SRCC$).
Figure \ref{fig:sim} illustrates the highest $\Delta SRCC$ recorded in 1000 simulation runs for each standard deviation of the Gaussian noise. Results show that SRCC is not robust to noise and (in the worst-case scenario) with small noise ($\mu=0$, $\sigma<0.1$) a large difference in Spearman's correlation coefficient can occur. 
The transformation leads to more robust coefficient calculation.

\section{Discussion and conclusion}

A subjective MOS value is a highly sensitive metric as votes given by all participants directly influence its numeric value. Even in repeatability studies, exact equal MOS values can rarely be observed.
Therefore MOS values are accompanied by 95\%~CIs.
We showed that conditions with small differences in MOS values (i.e. lower than their 95\%~CIs) are highly plausible within a dataset. 
Furthermore, we presented that applying rank-based statistics on MOS values might be misleading as those techniques consider two or more items to make a tied rank when their value is equivalent. 
We presented the maximum effect on SRCC, which depends on the number of missed tried ranks and number of conditions.
It should be noted that the effect of missed tied ranks approaches to zero when the number of conditions in the dataset increases.
Furthermore, we introduced a transformation method that considers two MOS values to make a tied rank when at least one of them lies in 95\%~CIs of the other one. 
Results of the simulation study showed that with the transformation method, the SRCC is less sensitive to white Gaussian noise.
We also recommend avoiding more than two fraction digits for MOS values when applying statistical techniques.
Equality of MOS values, as well as analyzing the effect of missed tied ranks on the non-parametric statistical test methods, are subjects of future work.

\bibliographystyle{IEEEtran}
\bibliography{library}

\end{document}